\documentclass[twocolumn,preprintnumbers,amsmath,amssymb]{revtex4}
\usepackage{graphicx}
\usepackage{dcolumn}
\usepackage{bm}
\begin{document}
\title{
Terahertz surface plasmons in optically pumped graphene 
structures 
}
\author{
A.~A.~Dubinov$^{1,2}$, V.~Ya.~Aleshkin$^2$, 
V.~Mitin$^3$,  T.~Otsuji$^{4,5}$,
and V.~Ryzhii$^{1,5}$
}

\affiliation{
$^1$Computational Nanoelectronics Laboratory,~University of Aizu, 
Aizu-Wakamatsu  965-8580, Japan\\ 
$^2$Institute for Physics of Microstructures, Russian Academy of Sciences,
Nizhny Novgorod 603950, Russia\\ 
$^3$Department of Electrical Engineering,
University at Buffalo, State University of New York, NY 14260, USA\\ 
$^4$Research Institute for Electrical Communication, Tohoku University, Sendai 980-8577, Japan\\
$^5$Japan Science and Technology Agency, CREST, Tokyo 107-0075, Japan
}
\date{\today}

\begin{abstract}
We analyze the surface plasmons (SPs) propagating along the optically pumped
single-graphene
layer (SGL) and multiple-graphene layer (MGL) structures. It is shown that at sufficiently strong optical pumping when the real part of dynamic conductivity 
of SGL and MGL structures
becomes negative in the terahertz (THz) range of frequencies
due to the interband population inversion, the damping of the THz
SPs can 
give way  to their amplification. This effect can be used in graphene-based THz lasers
and other devices. Due to relatively small SP group velocity, the absolute value of their absorption coefficient (SP gain) can be large, substantially exceeding that of the optically pumped structures with the dielectric waveguide. 
The comparison of the SGL and MGL structures  shows that to maximize the SP gain
the number of GL layers should be properly choosen.
\end{abstract}

\maketitle


\section{Introduction}
\vspace*{-0.5cm}
Optical excitation of graphene can result
in the interband population inversion~\cite{1,2}.
At sufficiently strong excitation, 
the interband emission of photons
can prevail over the intraband (Drude) absorption.
In this case, the real  part of the dynamic conductivity
Re~$\sigma_{\omega}$ 
becomes 
 negative. 
Due to the gapless energy spectrum of graphene~\cite{3},
Re~$\sigma_{\omega}$ can be negative at relatively low frequencies $\omega$,
for instance, those in the  terahertz (THz) range. 
This effect can be used in  graphene-based THz optically pumped
lasers with the Fabri-Perot resonators and the resonators
based on dielectric or slot-line waveguides~\cite{4,5,6}.
As was previously
pointed out~\cite{1} and analyzed~\cite{7},
apart from the lasing associated with the stimulated generation of electromagnetic
modes, the stimulated generation of
 different plasmons (with their conversion into 
electromagnetic radiation) can also be of practical interest.
The plasma excitations in graphene  were considered, in particular, 
in Refs.~\cite{8,9,10,11,12,13}.
In this paper, we study the propagation and amplification
of surface plasmons (SPs) in graphene structures
under nonequilibrium conditions when Re~$\sigma_{\omega} < 0$
due to optical pumping.
We consider a single-graphene layer (SGL) and a multiple-graphene layer (MGL)
on a substrate.  As demonstrated recently (see the review paper~\cite{14}
and the references therein), the MGLs, which constitute stacks of disoriented
(non-Bernal stacked) GLs, exhibit similar electron and optical
properties  as individual GLs, while the electron momentum relaxation
 time in such MGLs can be extremely long.
The latter circumstance implies that the intraband  absorption 
of photons and plasmons can be effectively diminished.
This makes the MGLs very prospective for different optoelectronic devices
considering that the net dynamic conductivity along GLs increases 
with increasing their number in the MGL structure.
The MGL structures can be particularly effective in lasers in which the THz modes
are supported by the external resonator or waveguide~\cite{5,6},
THz photodetectors~\cite{15,16,17}, and transit-time THz oscillators~\cite{18,19}.
However, SGL and MGL structures
can play the dual role: the imaginary  parts of their dynamic conductivity
provides the mode localization near the SGL or MGL (i.e., the 
formation of SPs), while the real part provide absorption or amplification
of plasmons. Due to this,
 the MGL structures as SP waveguides are 
not always superior over  the SGL 
structures.
\begin{figure}[t]
\vspace*{0.9cm}
\begin{center}
\includegraphics[width=7.5cm]{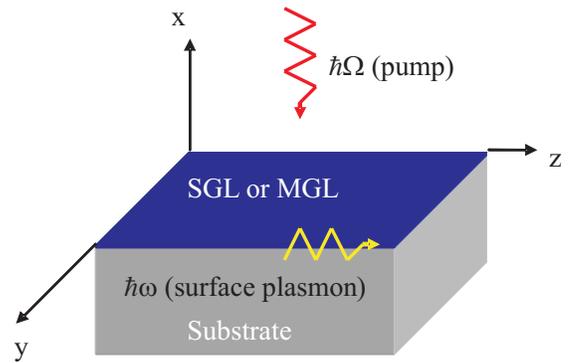}
\caption{Schematic view of the SGL/MGL structure under consideration.
}
\end{center}
\end{figure}
\begin{figure*}[t]
\vspace*{-0.4cm}
\begin{center}
\includegraphics[width=7.4cm]{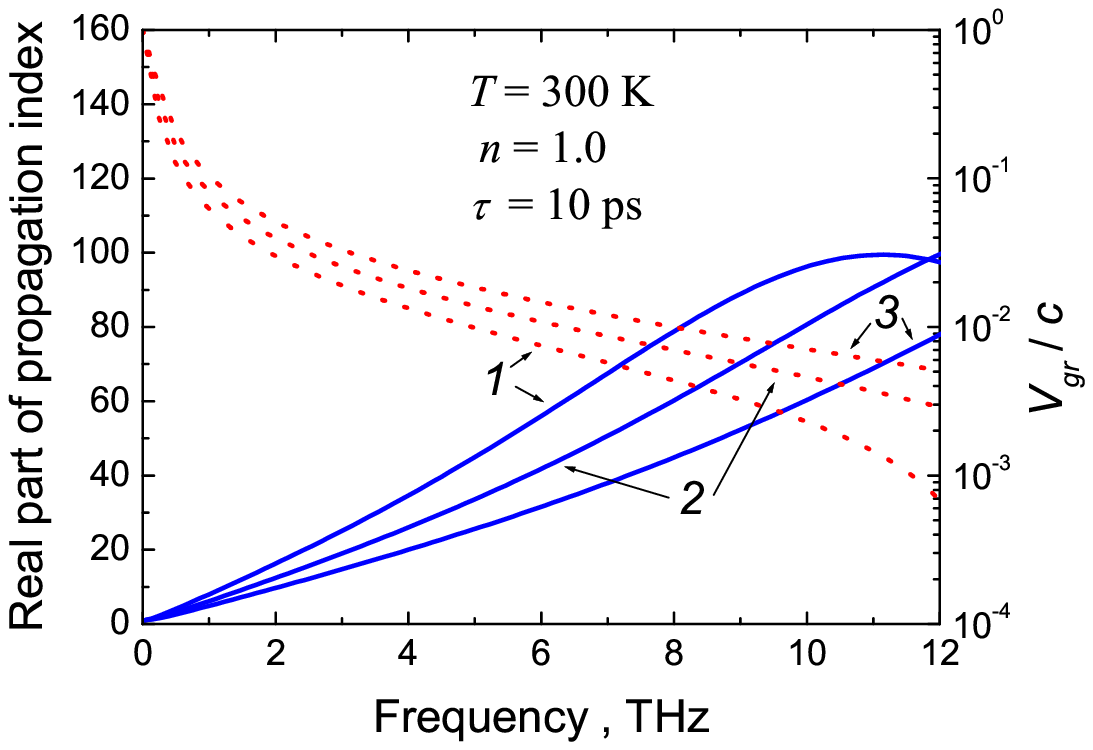}
\includegraphics[width=7.4cm]{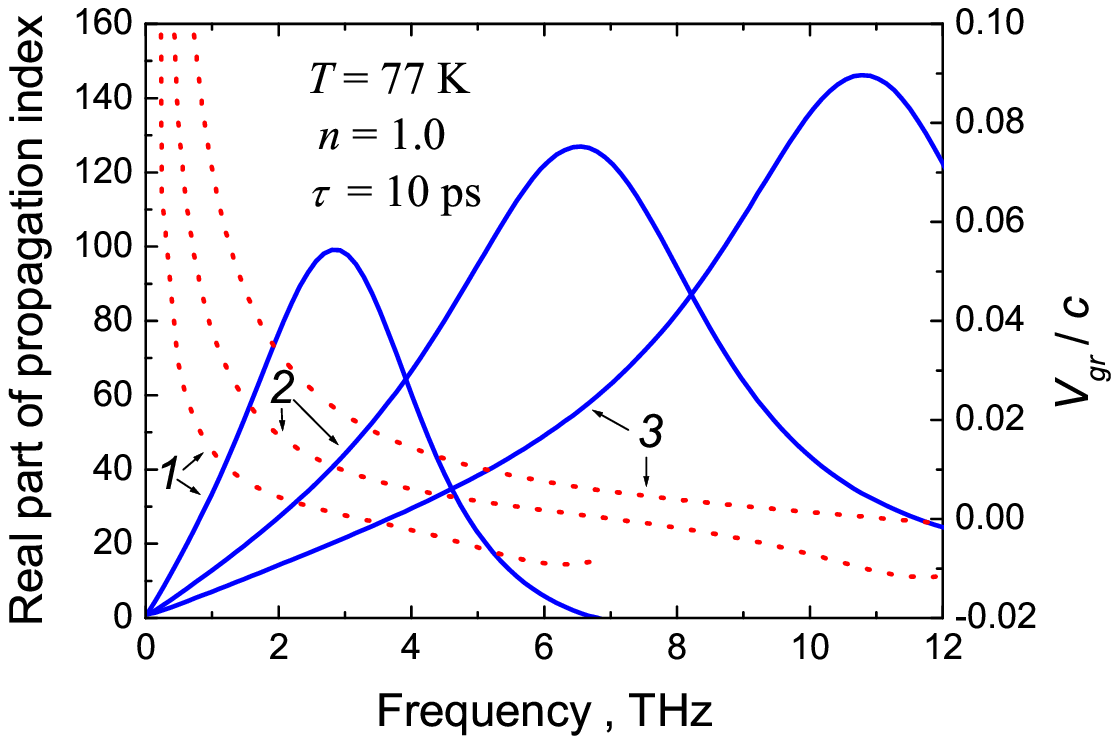}\\
\includegraphics[width=7.4cm]{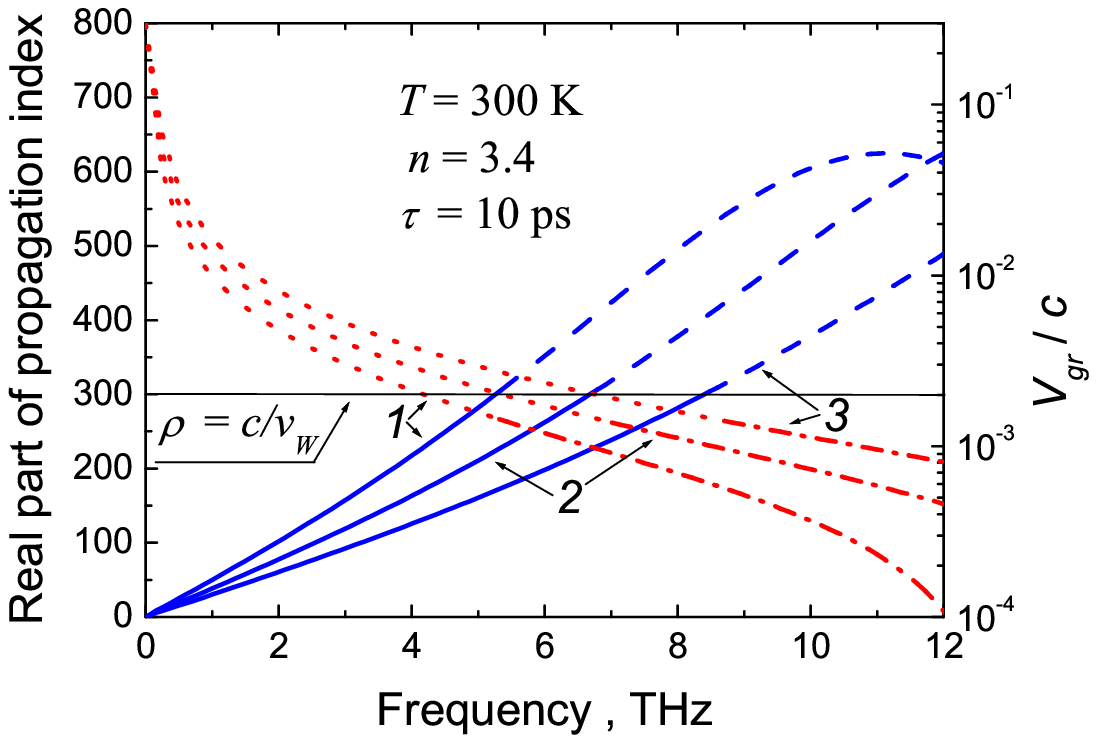}
\includegraphics[width=7.4cm]{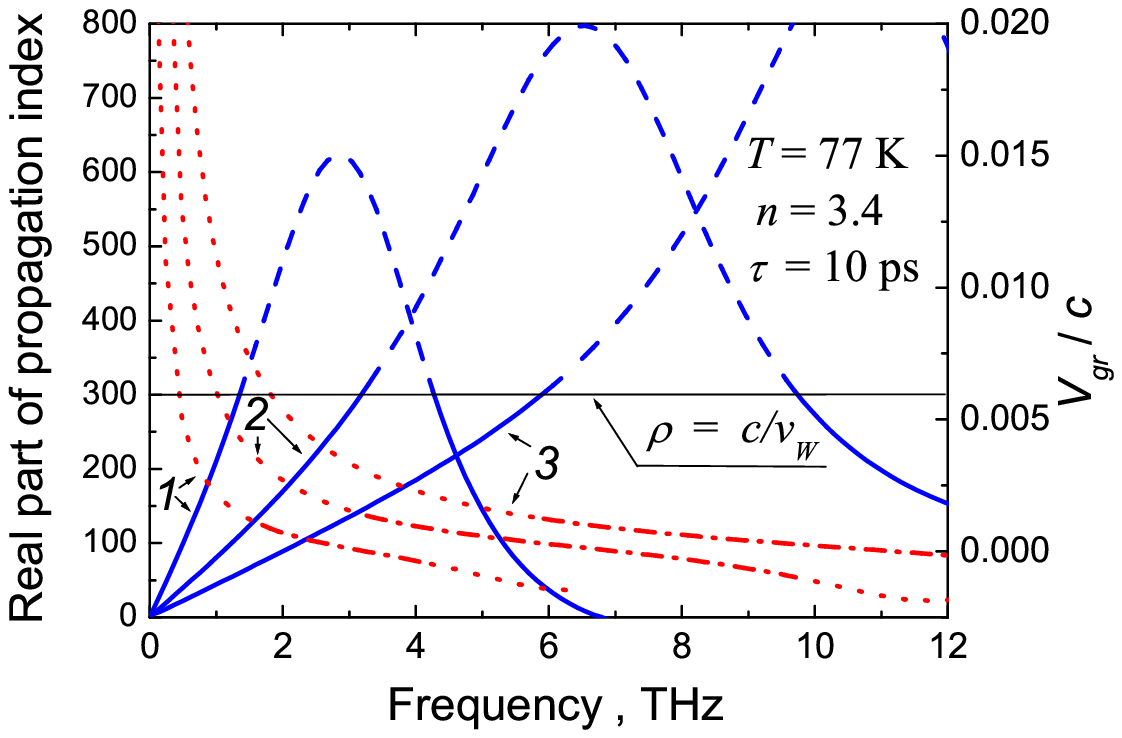}
\caption{Frequency dependence of the real part of propagation index 
Re~$(\rho)$ (solid lines)
and the SP group velocity normalized by the speed of light (dotted lines)
in SGL structures
at different temperatures ($T = 300$~K and $T = 77$~K) and
different quasi-Fermi energies  (1 -$\varepsilon_F^T = 0$~meV,   
2 -$\varepsilon_F^T = 10$~meV, and 3 -$\varepsilon_F^T = 20$~meV)
for SGL structures with different substrate refraction index ($n = 1.0$
and $n = 3.4$). Dashed and dashed-doted lines in figures for $n = 3.4$
correspond to the ranges
where the effects of spatial dispersion might be essential.
}
\end{center}
\end{figure*}    
Since the thickness, $d$, 
of the MGL structure with even relatively large number, $K$, of GLs is rather small in comparison with the wavelength of the plasmons
under consideration, we shall consider the SGL and MGL structure alike.
The real MGL usually include a highly
conducting  bottom GL (at the interface between the substrate
and other GLs) with rather high electron density~\cite{14,20}.
Although this bottom GL can be in some way removed,
we will take it into account.  

It is assumed that the GL or MGL structure is illuminated from the top
by light with the energy of photons $\hbar\Omega$.
Under optical excitation, the electron and hole densities exceed
substantially their equilibrium values. Due to this, one can
consider the electron and hole systems under consideration as characterized
by the quasi-Fermi energies $\pm\varepsilon_F^{(k)}$, respectively, 
 and the effective temperature $T$.
A distinction in the Fermi energies in GLs with different indices is due
to the attenuation of incident optical radiation associated
with its absorption in the GLs closer to the structure top: 
$\varepsilon_F^{(k)} < \varepsilon_F^{(K)} = \varepsilon_F^{T}$.
The electron-hole system in the bottom GL in MGLs 
  is characterized by
the unified Fermi energy $\varepsilon_F^B$ determined by the doping
of this GL.
If the characteristic time, $\tau_0$, of the emission of the optical phonon 
by an electron or a hole is much shorter than the time of pair collisions,
the photoexcited electrons and holes emit a cascade of optical phonons
and occupy low energy states in the conduction and valence bands, respectively.
In this case, the contribution of optical excitation to the heating
of the electron-hole system is small, so that the effective 
temperature $T$ is close to the lattice temperature $T_l$~\cite{21}.
In the opposite case, the photoexcited electrons and holes are thermalized,
and their effective temperature is determined by the pumping photon
energy $\hbar\Omega$ and the rate of electron and hole
 energy relaxation.
 In such a situation, the effective temperature
can be elevated, so that 
somewhat  stronger optical pumping might be needed to fullfil
the condition Re~$\sigma_{\omega} < 0$ ~\cite{22}. 
\vspace*{0.5cm}
\section{Dynamic conductivity of SGL and MGL structures}

The net dynamic conductivity in the lateral direction of a  structure with
$K$ GLs at the signal frequency $\omega$
can be  presented as the sum of the contributions 
of the individual GLs 
$\sigma_{\omega}^{(k)}$  ($k = 1, 2,..., K$) 
and 
the bottom GL  $\sigma_{\omega}^{B}$;

\begin{equation}\label{eq1}
\sigma_{\omega}
= \sum_{k=1}^K\sigma_{\omega}^{(k)} +  \sigma_{\omega}^{B}.
\end{equation}
Considering the expressions for $\sigma_{\omega}^{(k)}$ 
and  $\sigma_{\omega}^{B}$   obtained previously (see, for instance,
Refs,~\cite{10,23}),
one can arrive at the following:
\begin{widetext}
\begin{equation}
\sigma_{\omega}^{(k)} = 
\displaystyle\biggl(\frac{e^2}{4\hbar}\biggr)\biggl\{
\frac{8k_BT\tau}
{\pi\hbar(1 -  i\omega^2\tau)}
\ln \biggl[1 + \exp\biggl(\frac{\varepsilon_F^{(k)}}{k_BT}\biggr)\biggr]
+
 \displaystyle
\tanh\biggl(\frac{\hbar\omega - 2\varepsilon_F^{(k)}}{4k_BT}\biggr)
-  \frac{4\hbar\omega}{i\pi}\int_0^{\infty}\frac{G(\varepsilon, \varepsilon_F^{k})- G(\hbar\omega/2, \varepsilon_F^{k}) }{(\hbar\omega)^2 - 4\varepsilon^2}d\varepsilon\biggr\},
\end{equation}
$$
\sigma_{\omega}^{B} = 
\displaystyle
\biggl(\frac{e^2}{4\hbar}\biggr)
\displaystyle
\frac{4k_BT\tau_B}
{\pi\hbar(1 - i\omega\tau_B)}
\ln \biggl[1 + \exp\biggl(\frac{\varepsilon_F^{B}}{k_BT}\biggr)\biggr]
$$
\begin{equation}\label{eq3}
+ \displaystyle
\biggl(\frac{e^2}{4\hbar}\biggr)
\biggl\{1 - \biggl[1 + \exp\biggl(\frac{\hbar\omega/2 - \varepsilon_F^B}{k_BT}\biggr)\biggr]^{-1} - \biggl[1 + \exp\biggl(\frac{\hbar\omega/2 + \varepsilon_F^B}{k_BT}\biggr)\biggr]^{-1}
- \frac{4\hbar\omega}{i\pi}\int_0^{\infty}\frac{G(\varepsilon, \varepsilon_F^B)- G(\hbar\omega/2, \varepsilon_F^B) }{(\hbar\omega)^2 - 
4\varepsilon^2}d\varepsilon\biggr\}.
\end{equation}
\end{widetext}
Here, $e$ is the electron charge,  $\tau_B$ and  $\tau$ are the electron and hole momentum relaxation 
times in the bottom and other GLs, respectively, $\hbar$ is the reduced Planck constant,
$k_B$ is the Boltzmann constant, and
\begin{equation}\label{eq4}
G(\varepsilon, \varepsilon^{\prime}) = \frac{\sinh(\varepsilon/k_BT)}
{\cosh(\varepsilon/k_BT) + \cosh(\varepsilon^{\prime}/k_BT)}.
\end{equation}
In the case of SGL structure, one should put $K = 1$ and $\sigma_{\omega}^B = 0$.
For the 
MGL structures without the bottom GL, one can use Eq.~(3) 
formally setting 
$\sigma_{\omega}^B = 0$.

The quasi-Fermi energies in the GLs with $k \geq 1$,  $\varepsilon_F^{(k)}$,
 are mainly determined
by the electron (hole) density in  this layer $\Sigma_k$,
with  
 $\varepsilon_F^{(k)}  \propto \sqrt{\Sigma^{(k)}}$ (at sufficiently strong degeneracy
of the electron and hole systems), i.e.,  
by the rate of photogeneration  $G_{\Omega}^{(k)}$ 
by the optical
radiation  at the $k-$th GL plane.
Considering the attenuation of the optical 
pumping radiation due to its absorption
in each GL, one can obtain that
  $\varepsilon_F^{(k)}$ can be expressed via
the quasi-Fermi energy
in the topmost GL
$\varepsilon_F^{T} = \varepsilon_F^{(K)}$, which, in turn, is a function of
the  the intensity of incident pumping radiation $I_{\Omega}$.
Considering the MGL structures, the
 $\varepsilon_F^{(k)}$ versus $\varepsilon_F^{T}$ dependence was found as
previously~\cite{6}.

\section{Dispersion equation for SP.}

\vspace*{-0.4cm}

\begin{figure}[b]
\vspace*{-0.4cm}
\begin{center}
\includegraphics[width=7.5cm]{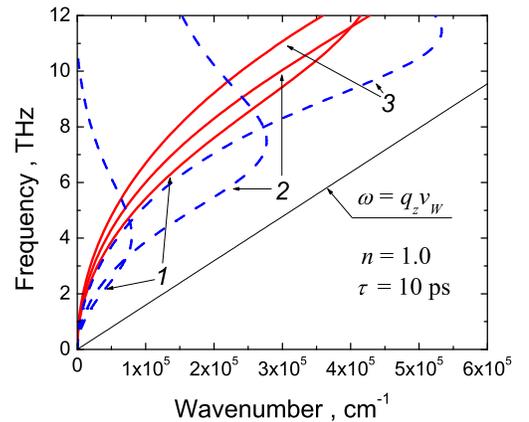}
\caption{Dispersion of SPs 
in a SGL structure  at $T = 300$~K (solid lines) and $T = 77$~K (dashed lines)
with the same as in Fig.~2 values of the quasi-Fermi energy.
}
\end{center}
\end{figure}    

\begin{figure}[t]
\vspace*{-0.4cm}
\begin{center}
\includegraphics[width=7.5cm]{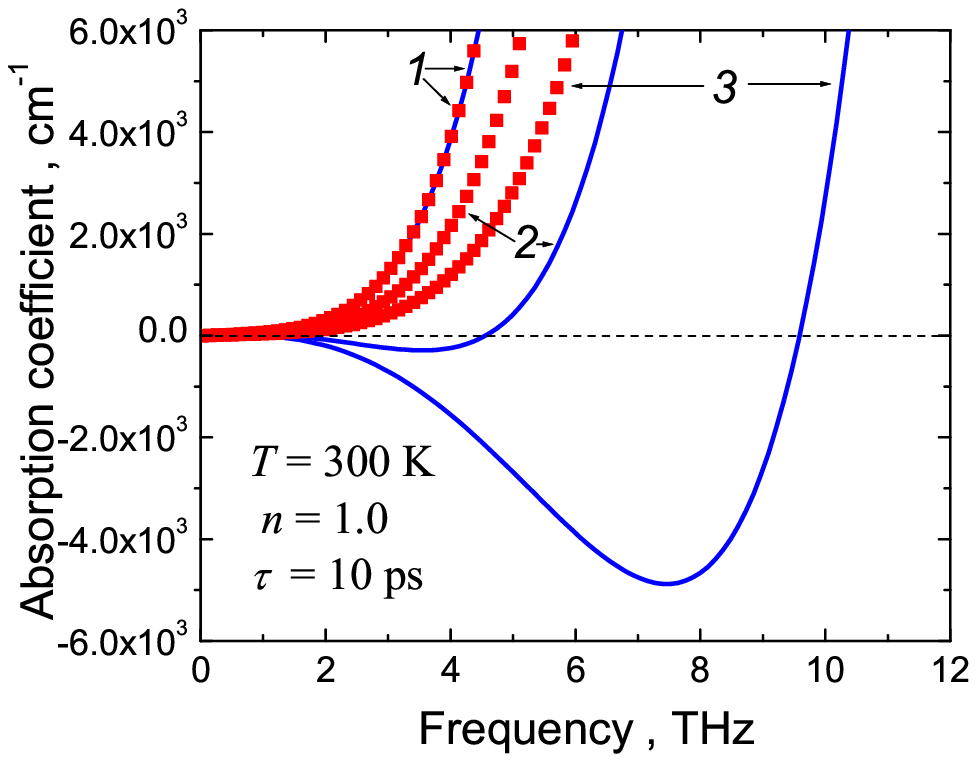}\\
\includegraphics[width=7.5cm]{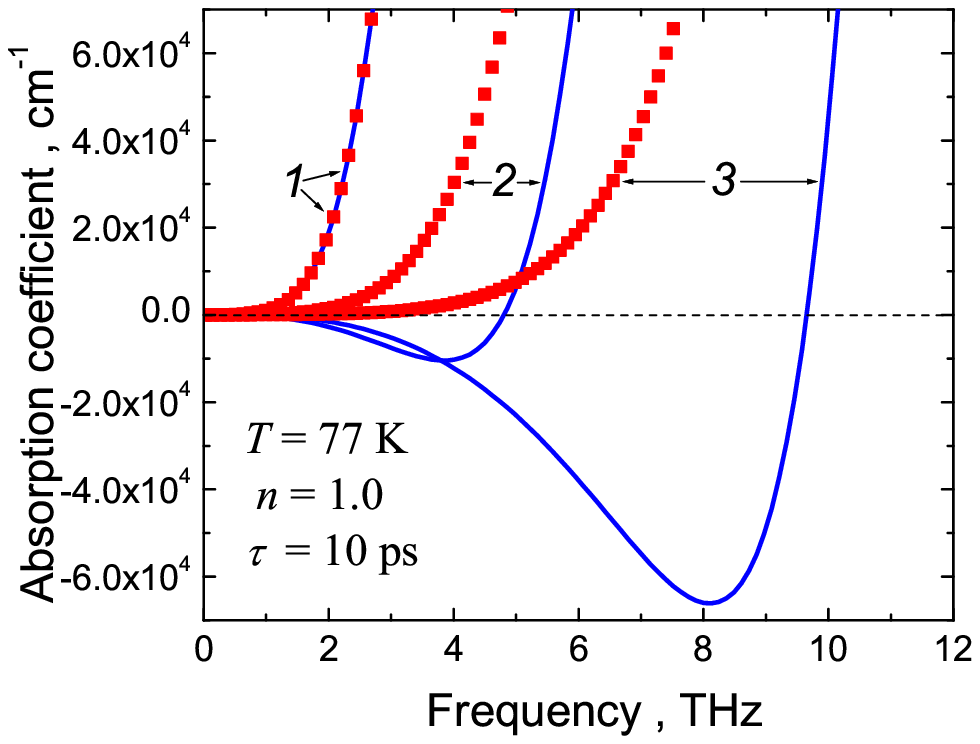}
\caption{Frequency dependences of SP absorption coefficient
2Im~$(q)$ in a SGL structure
at $T = 300$~K (upper panel)  and $T = 77$~K (lowe panel) and different 
quasi-Fermi energies (1 -$\varepsilon_F^T = 0$~meV,   
2 -$\varepsilon_F^T = 10$~meV, and 3 -$\varepsilon_F^T = 20$~meV).
Markers show the dependences for equilibrium electron-hole systems calculated using 
the derived formulas as well as 
using~\cite{12} for intrinsic
and doped SGL structures.
}
\end{center}
\end{figure} 
\begin{figure}[t]
\vspace*{-0.4cm}
\begin{center}
\includegraphics[width=7.5cm]{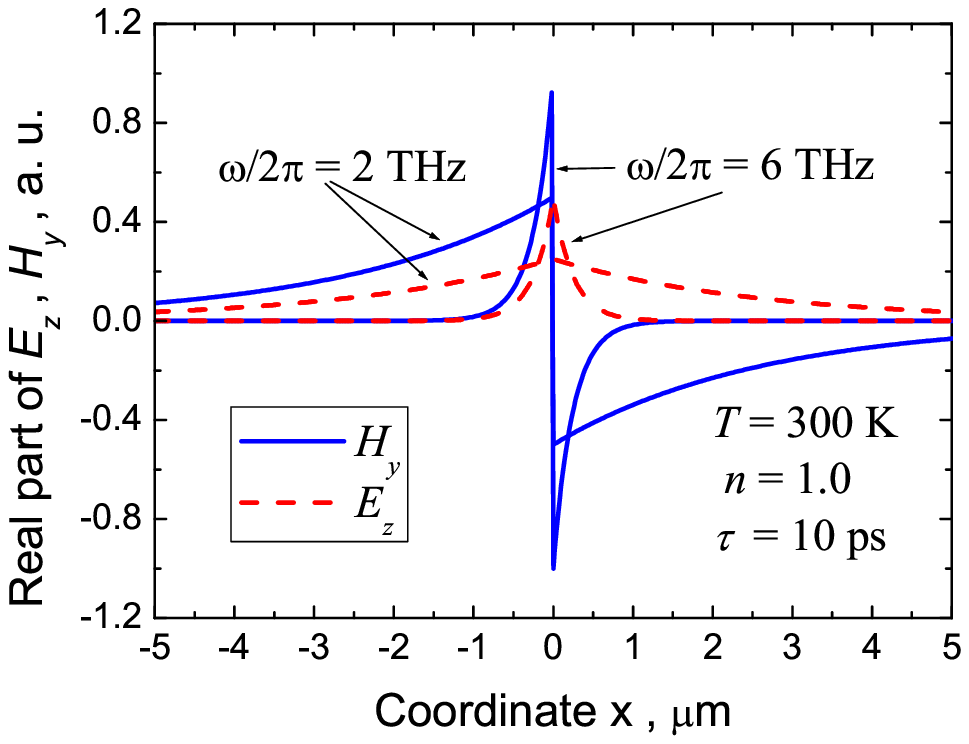}\\
\includegraphics[width=7.5cm]{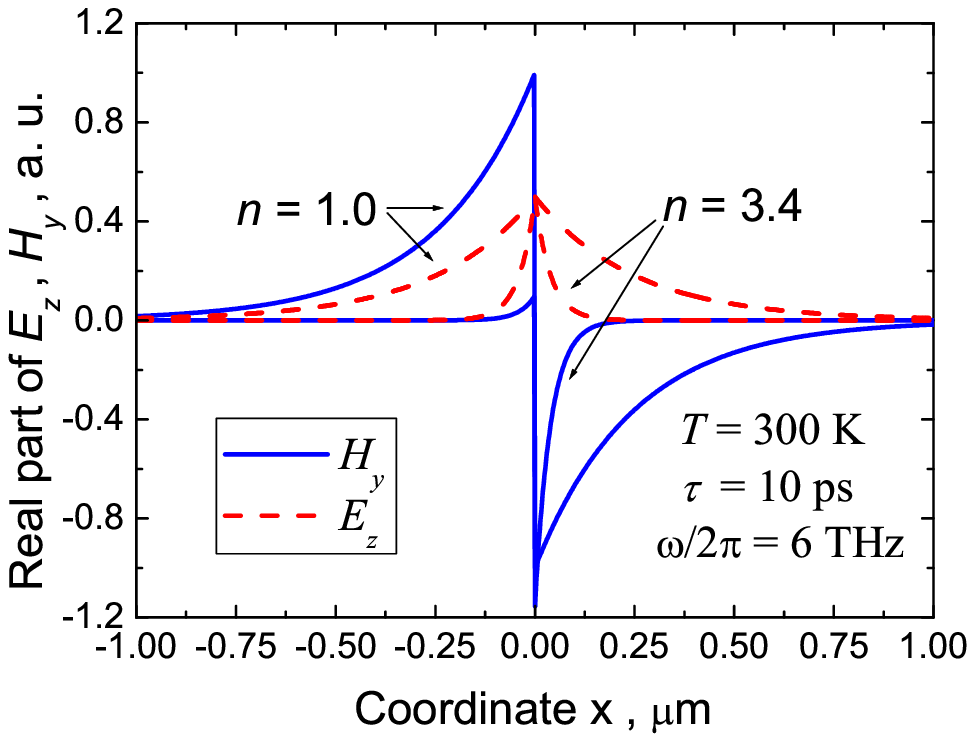}
\caption{
Spatial distributions of the real parts of electric and magnetic fields 
in SGL structures at different frequencies (upper panel)
and different substrate refraction indices (lower panel): $T = 300$~K.
$\varepsilon_F^T = 20$~meV, and $\tau = 10$~ps.
}
\end{center}
\end{figure}  

Using the the Maxwell equations, in which
the current associated with electron-hole system is given by
of 
\begin{equation}\label{eq5}
j_x = 0, \qquad
j_y = 0, \qquad
j_z = \sigma_{\omega}E_z\,\delta(x),
\end{equation}
The components of the electric and magnetic fields, ${\bf E} = (E_x, 0, E_z)$
and ${\bf H}= (0, H_y, 0)$ can be presented in the following form:
\begin{equation}\label{eq6}
E_x  \propto E_z \propto  H_y \propto
\exp\biggl(i\frac{\omega}{c}\sqrt{1 - \rho^2}\,x + i\frac{\omega}{c}\rho\,z - i\omega\,t\biggr), 
\end{equation}
at $x \geq 0$, and
\begin{equation}\label{eq7}
E_x  \propto E_z \propto  H_y \propto
\exp\biggl(-i\frac{\omega}{c}\sqrt{n^2 - \rho^2}\,x + i\frac{\omega}{c}\rho\,z - i\omega\,t\biggr) 
\end{equation}
at $x < 0$.
Here, $\omega$ is the signal (plasmon) frequency, $c$ is the speed of light,
$n$ is the refraction index (generally complex) of the substrate,
$\rho$ is the complex propagation index (which relates to the wavenumber 
$q_z = \rho\omega/c$),
the axis $x$ is directed perpendicular to the structure plane,
the axis $z$ corresponds to the in-plane direction along the SP
propagation, and $\delta (x)$ is the Dirac delta function. The latter
 describes
the localization of the plasmon near the SGL or MGL structure plane.
Considering Eqs.~(6) and (7), from the Maxwell equations we obtain
the following dispersion equation for SPs: 
\begin{equation}\label{eq8}
\sqrt{n^2 - \rho^2} + n^2\sqrt{1 - \rho^2} + \frac{4\pi}{c}\sigma_{\omega}\sqrt{1 - \rho^2}\sqrt{n^2 - \rho^2} = 0.
\end{equation}
Equation~(8) can be derived from  that obtained previously using a different
approach~\cite{12,13}.
What is important for the present consideration is that $\sigma_{\omega}$
is determined by both nonequilibrium  electron and hole components
with, in particular, Re~$(\sigma_{\omega}) < 0$.
Solving Eq.~(8) with respect to $\rho$, one can find
the propagation index Re~$(\rho)$ and the SP absorption  (amplification)
coefficient 2Im~$(q) = 2{\rm Im}(\rho\omega/c)$ or the SP gain -2Im~ $(q)$.
In the cases when $n = 1$ (SGL or MGL structure levitating in the free space),
Eq.~(8) obviously yields

\begin{equation}\label{eq9}
\rho = \sqrt{1 - \frac{c^2}{4\pi^2\sigma_{\omega}^2}}. 
\end{equation}

\section{Results of calculations}
\vspace*{-0.4cm}

\subsection{SP spectra and SP damping/amplification}
\vspace*{-0.4cm}

Figure~2 presents  the frequency dependences of the real part of the propagation index Re~$(\rho) \propto {\rm Re}(q)$ shown by solid lines
calculated for SGL structures with different substrate refraction indices $n$
at different temperatures $T$ and different values of the quasi-Fermi energies 
$\varepsilon_F^T$  (i.e., for different
pumping intensities).
Dotted lines in Fig.~2 correspond to the frequency dependences of the group velocity
of SPs normalized by the speed of light in vacuum $c$.
In our model we disregarded the effect of spatial dispersion on the dynamic conductivity of SGL and MGL structures, i.e., the dependence of $\sigma_{\omega}$
on the SP wavenumber $q_z$ in Eqs.~(1) - (4). 
This is valid if $\rho < c/v_W$
or $\omega < {\rm Re} (q_z)v_W$~\cite{10}, where $v_W = 10^8$~cm/s is the characteristic velocity of electrons in holes in graphene. In the case of SGL structures with
$n = 1$, these inequalities are satisfied in the frequency range under consideration (see Figs.~2 and 3). However, as follows
 from Fig.~2 at larger $n$, the validity of the dependences obtained
is limited by relatively low frequencies 
($\omega/2\pi \lesssim 5 - 8$~THz for $n = 3.4$ at $T = 300$~K).
One can see from Fig.~2 that the SP group velocity can be negative (at $T = 77$~K)
at elevated frequencies. This corresponds to backward waves~\cite{24,25}
and  occurs at the frequencies at which the imaginary
part of the dynamic conductivity is determened primarily by the interband transitions.
Figure~3 shows the spectra of SPs, i.e.,
$\omega$ versus Re$(q_z)$ dependences in the SGL structures at 
different temperatures
at the same pumping conditions as in Fig.~2.


Figure~4 shows the SP absorption coefficient 2Im~$(q_z)$
as a function of frequency calculated for different temperatures $T$
and different values of the quasi-Fermi energies 
$\varepsilon_F^T$. One can see that an increase in
$\varepsilon_F^T$ leads to  widening of the frequency range where  Im~$(q_z) < 0$
and a marked increase in the absolute
value of the absorption coefficient in this range.
The dependences indicated by markers were calculated for  equilibrium
SGLs using the above formulas which  in the absence of optical pumping
at  $n = 1$ coincide with those obtained previously~\cite{12}.
The markers correspond to  intrinsic  (with the Fermi energy $\varepsilon_F = 0$) and doped SGL structures.
The  dependences for doped GL structures at the equilibrium
 correspond to the Fermi
energies $\varepsilon_F$
in doped SGLs at  the equilibrium equal to the quasi-Fermi energies 
$\varepsilon_F^T$ in the optically pumped SGLs. Clear distinctions
in these dependences in the equilibrium and pumped GLs 
are associated with the contributions of both
electrons and holes to the negative dynamic conductivity
due to the interband transitions (in the latter). 

Figure~5 shows the spatial distributions (in the direction perpendicular to the SGL plane) of the SP electric and magnetic fields in the SGL structures with 
different substrate
refraction indices.
\begin{figure}[t]
\vspace*{-0.4cm}
\begin{center}
\includegraphics[width=7.4cm]{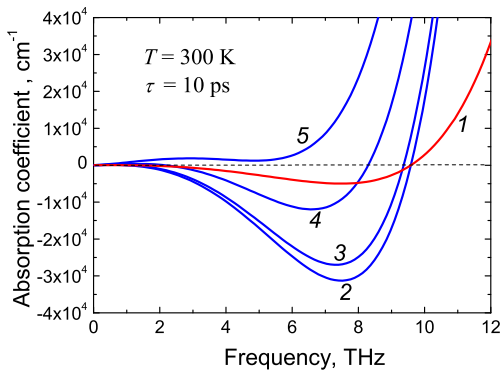}
\caption{Frequency dependences of SP absorption for SGL structures
with  different  substrate
refraction indices $n$:
1 - $n = 1.0$, $2 - n = 3.4$, $3 - n = 3.4 + i0.01$, and  
$4 - n = 3.4 + i0.05$, and  $5 - n = 3.4 + i0.1$
($T = 300$~K, $\varepsilon_F^{T} = 20$~meV).
}
\end{center}
\end{figure}  

\begin{figure}[h]
\vspace*{-0.4cm}
\begin{center}
\includegraphics[width=7.4cm]{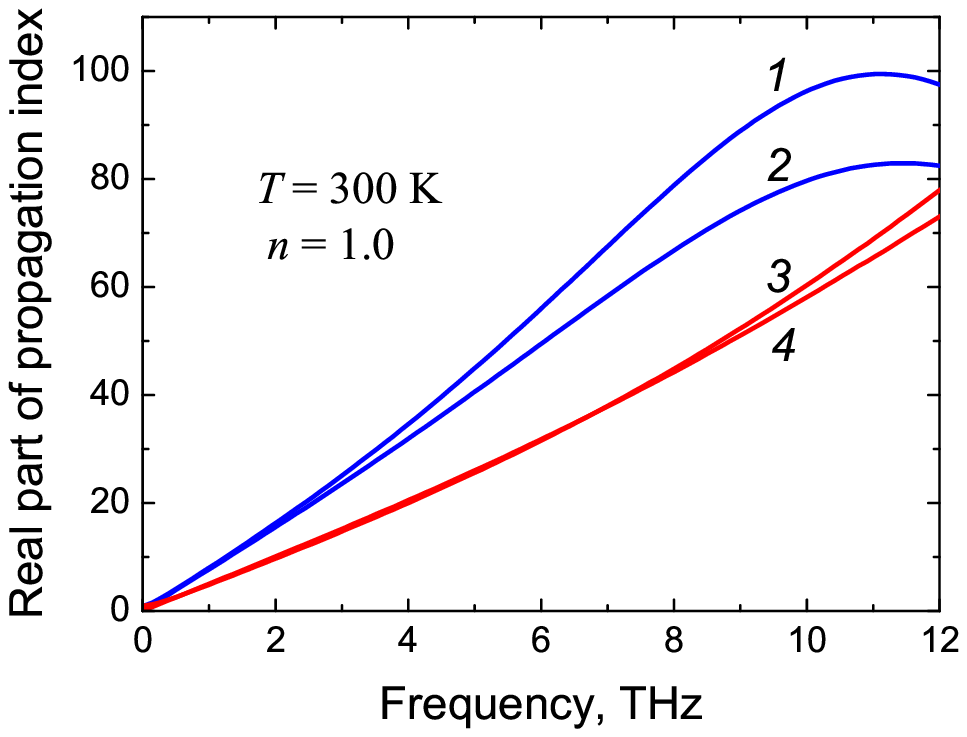}\\
\includegraphics[width=7.7cm]{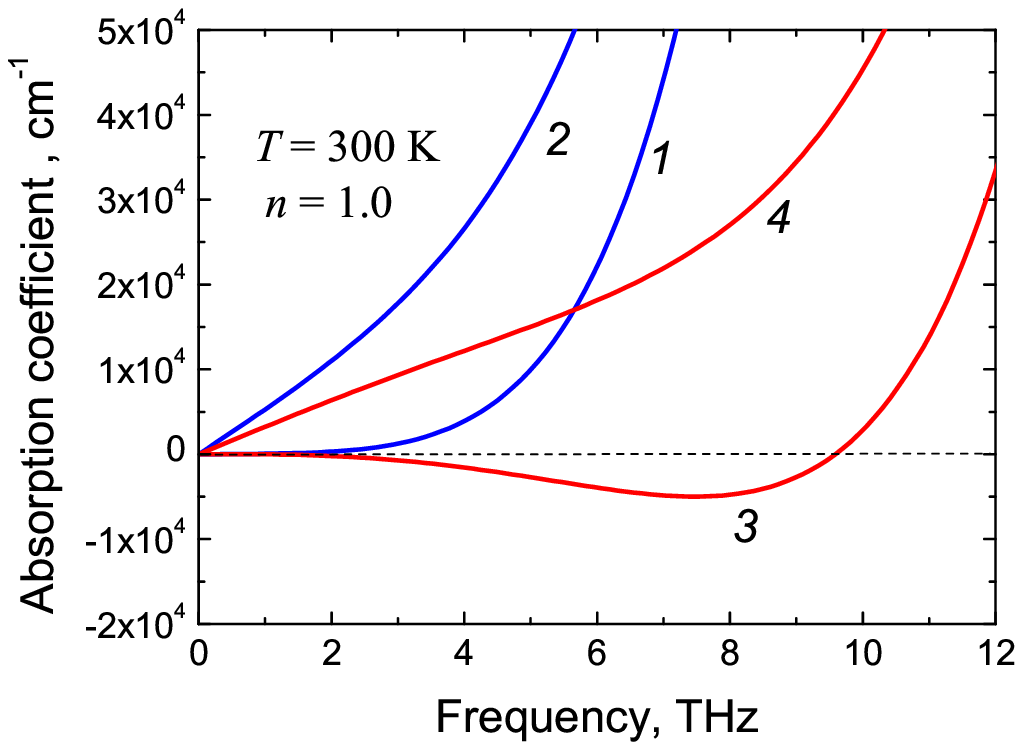}
\caption{ Real part of SP 
propagation index (upper panel) and absorption coefficient (lower panel) in SGL structures
as functions of frequency: 
$1 - \varepsilon_F^T = 0$~meV, $\tau = 10$~ps ,
$2 - \varepsilon_F^T = 0$~meV, $\tau = 0.1$~ps,
$3 - \varepsilon_F^T = 20$~meV, $\tau = 10$~ps, and 
$4 - \varepsilon_F^T = 20$~meV, $\tau = 0.1$~ps.
}
\end{center}
\end{figure}

\begin{figure}[t]
\vspace*{-0.4cm}
\begin{center}
\includegraphics[width=7.4cm]{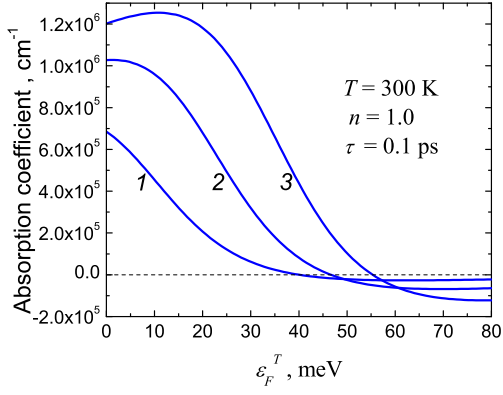}
\caption{SP absorption coefficient as a function of the quasi-Fermi energy
in SGL structures with short momentum relaxation time ($\tau = 0.1$~ps)
at $1 - \omega/2\pi = 15$~THz,  $2 - \omega/2\pi = 20$~THz, 
$3 - \omega/2\pi = 25$~THz.
}
\end{center}
\end{figure}

\subsection{Role of substrate and intraband absorption}
\vspace*{-0.4cm}
Figure~6 shows the frequency dependences of the SP absorption coefficient 
calculated for SGLs with the substrates with different real and imaginary parts of 
the refraction index (different absorption in the substrate).
As follows from Fig.~6, the contribution of the substrate to the SP absorption
can be insignificant at 
 the realistic values of the  imaginary part of the substrate refraction index.
In particular, in the case of the substrate made of undoped Si 
(Im~$(n) \simeq 3\times10^{-4}$~\cite{26}), the  imaginary part of 
the refraction index can be smaller than those used in the calculations of curves 3- 5
in Fig.~6.

Figure~7 shows the frequency dependences of the real part of the propagation
index and the SP absorption coefficient. These dependences were calculated for SGL structures with different electron and hole
 momentum relaxation times ($\tau = 10$~ps and $\tau = 0.1$~ps) 
for different $\varepsilon_F^T$,
i.e., for different pumping intensities assuming that
$T = 300$~K. As seen from Fig.~7, in the SGL structure 
with a relatively long momentum relaxation
time ($\tau = 10$~ps), the SP absorption coefficient changes its sign at moderate values of $\varepsilon_F^T$ ( $\varepsilon_F^T \lesssim 20$~meV).
However when  $\tau = 0.1$~ps, the SP absorption coefficient does not change 
the sign at least at $\varepsilon_F^T \sim 20$~meV (see below), although
it is markedly smaller than in the equilibrium conditions (without pumping).
However, at elevated pumping intensities (elevated values of the quasi-Fermi energy) 
the absorption coefficient 
in the SLG structures even 
with short momentum relaxation times can become negative although at higher frequencies
(see Fig.~8).


\begin{figure}[t]
\vspace*{-0.4cm}
\begin{center}
\includegraphics[width=7.9cm]{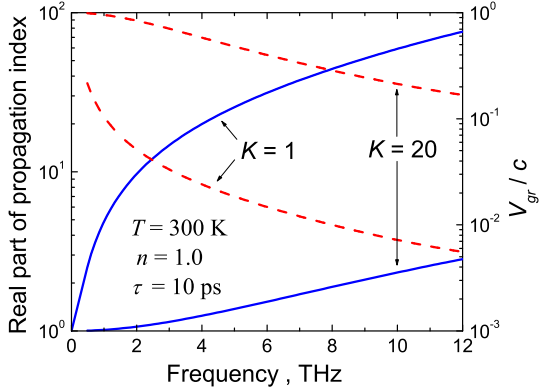}
\caption{Frequency dependences of 
real part of SP propagation index (solid lines)
and normalized group velocity (dashed lines) 
 for   SGL structure ($K = 1$)
and MGL structure ($K = 20$) at
$\varepsilon_F^T = 20$~meV.
}
\end{center}
\end{figure}  

\begin{figure}[t]
\vspace*{-0.4cm}
\begin{center}
\includegraphics[width=7.0cm]{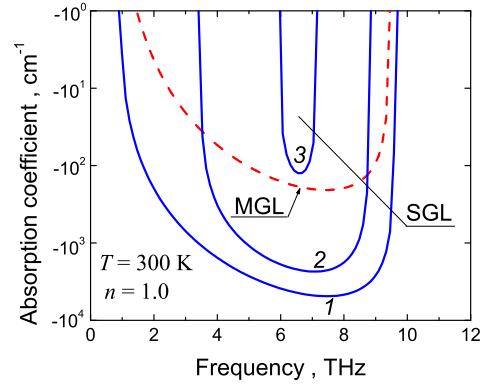}
\caption{Frequency dependences of  SP absorption coefficient at
$\varepsilon_F^T = 20$~meV
 for    SGL structure  with 1- $\tau = 10$~ps, 2-  $\tau = 1.0$~ps, and 
3-  $\tau = 0.54$~ps    and  for  MGL structure  with $K = 20$ (dashed line)
and $\tau = 10$~ps.
}
\end{center}
\end{figure}  
\begin{figure}[t]
\vspace*{-0.4cm}
\begin{center}
\includegraphics[width=7.5cm]{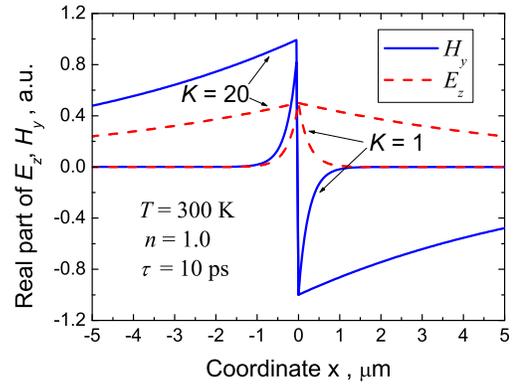}
\caption{Spatial distribution of real parts of the SP electric and magnetic fields
in a SGL  structure  and MGL structures with $K = 20$
 at $T = 300$~K, 
$\varepsilon_F^{T} = 20$~meV,  $\tau = 10$~ps, and $\omega/2\pi = 6$~THz.
}
\end{center}
\end{figure}

\begin{figure}[t]
\vspace*{-0.4cm}
\begin{center}
\includegraphics[width=7.4cm]{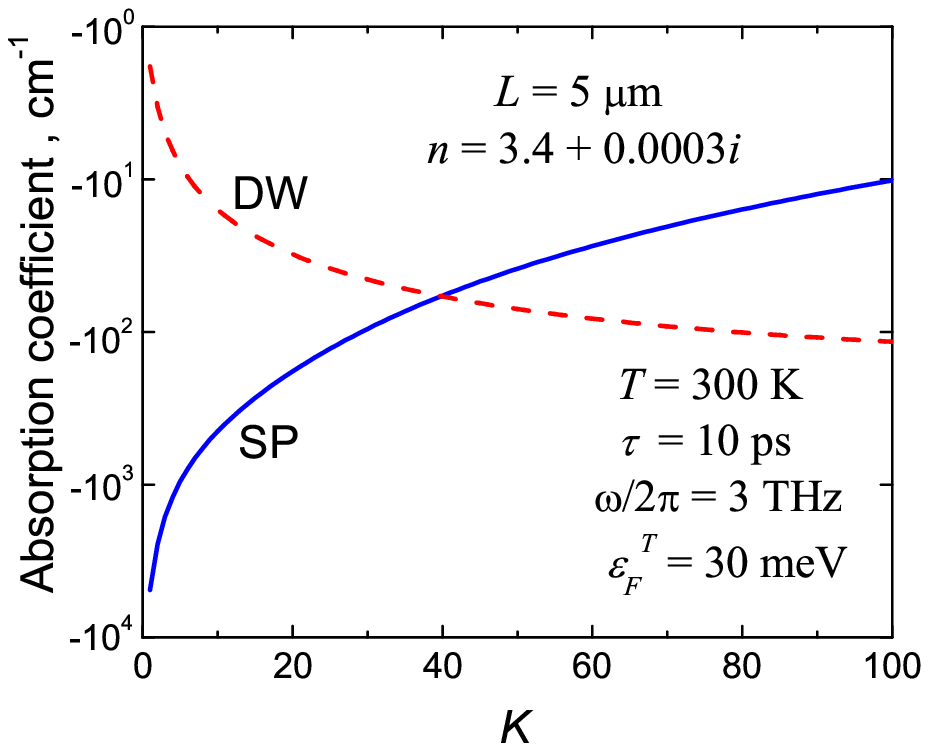}
\caption{Absorption coefficients of
SPs (solid line) and an electromagnetic mode (dashed line)
in a  dielectric waveguide (DW)~\cite{6} as functions of the number of GLs $K$.
}
\end{center}
\end{figure}

\subsection{Comparison of SGLs and MGLs}

\vspace*{-0.4cm}
In Figs.~9 and 10, we compare the real parts of the SP propagation index,
absorption coefficients, 
 the SP group velocities in the optically pumped SGLs and MGLs ($\varepsilon_F^T = 20$~meV). As seen, the SP characteristics in SGLs and MGLs are markedly different. In particular, SGLs exhibit substantially
stronger (about 50 times in the peaks) amplification (higher plasmon gain) in a wide 
frequencies range (from 4 to 8 THz) at $\tau = 10$~ps.
This is attributed to markedly different 
values of the SP group velocities 
in SGLs (relatively small plasmon group velocity)
and in MGLs (in which this velocity
is  relatively high), as shown in Fig.~9, which, in turn, is due to different  net electron and hole densities and 
different widths of the SP  electric field localization that is
clearly seen in Fig.~11. An example of 
the dependence of the SP absorption 
coefficient in the structures with optically pumped GLs on the number of the latter
is demonstrated in Fig.~12.
Thus, at the same values of the optical pumping intensity
and the electron and hole momentum relaxation time,
SPs in SGL structures  can exhibit stronger amplification than MGL structures.
However, as demonstrated~\cite{14}, the momentum relaxation time in epitaxially grown
MGLs can be rather
long, so MGL structures might be preferable in applications. 
Taking into account the real possiblity of  a large difference
in the electron and hole relaxation times in SGL and MGL structures,
in Fig.~10, we demonstrate that the SP amplification in the 
MGL structure with $K = 20$ and $\tau = 10$~ps can exceed that in the SGL structure
with smaller $\tau$. 
Hence, to achieve the SP maximum
gain, 
the number of GL layers should be optimal.

A decrease  in the SP gain with increasing $K$ is in contrast
to the behavior the electromagnetic modes in optically pumped MGL structures 
with the dielectric waveguide  considered 
by us previously~\cite{6}.
For comparison, the pertinent dependence taken from ~Ref.~\cite{6} for
 the waveguide thickness $L = 5~\mu$m
(at the same quality of GLs and pumping conditions) is shown in Fig.~12
as well.

\section{Conclusions}
\vspace*{-0.4cm}
We studied the SPs in optically pumped SGL and MGL structures.
Using the developed model, we calculated the SP  dispersion relations, spatial distributions of their electric and magnetic fields, and frequency dependences of
the absorption coefficient
as functions of the optical pumping  and factors determining the 
intraband absorption in GLs and the substrate.
It was demonstrated that at sufficiently strong but realistic optical pumping,
the SP absorption coefficient can be negative 
(so that the gain is positive) in the certain range of frequencies. 
The absolute value of the SP absorption coefficient (plasmon gain)
in SGL structure can be fairly large and
markedly exceed the gain of electromagnetic modes in
the dielectric waveguides with optically pumped SGL or MGL structures.  
In contrast to the cases
of amplification of electromagnetic modes
in the structures
with optically pumped SGL and MGL and dielectric waveguide,
the SP amplification weakens  with increasing number of GLs $K$.
However, due to possibility of rather long momentum relaxation times of electrons and holes in MGL structures,
the SP maximum  gain can be achieved at optimal values of the nimber of GLs.
The effect of SP amplification in optically pumped SGL and MGL structures
can be used in THz lasers (with the conversion of SPs to
the output electromagnetic modes). The variation of the SP
 characteristics by optical pumping  might be useful in different
GL-based plasma-wave devices. 

\vspace*{0.3cm}
\section*{Acknowledgments}
\vspace*{-0.6cm}
The authors are grateful to 
M.~Ryzhii
for comments.
This work was supported by the Japan Science and Technology Agency, CREST, 
Japan and partially by the Federal Russian Program 
"Scientific and Educational Staff", Russia.

\end{document}